\documentstyle[prd,aps]{revtex}
\begin{document}
\setlength{\unitlength}{1cm}
\thispagestyle{empty}
\null
\hfill WUE-ITP-97.026\\
\null
\hfill UWThPh-1997-26\\
\null
\hfill hep-ph/9708481\\
\vskip .8cm
\begin{center}
{\Large \bf Spin Correlations in Production\\[.5em]
and Subsequent Decay of Neutralinos%
\footnote{Work supported by the German Federal Ministry for
Research and Technology (BMBF) under contract number
05 7WZ91P (0).}
}
\vskip 2.5em
{\large
{\sc G. Moortgat-Pick}%
\footnote{e-mail: gudi@physik.uni-wuerzburg.de}
}\\[1ex]
{\normalsize \it
Institut f\"ur Theoretische Physik, Universit\"at W\"urzburg, Am
Hubland, D-97074~W\"urzburg, Germany,\\
Institut f\"ur Theoretische Physik, Universit\"at Wien,
Boltzmanngasse 5,\\
A-1090 Wien, Austria}\\[2ex]
{\large
{\sc H. Fraas}%
\footnote{e-mail: fraas@physik.uni-wuerzburg.de}
}\\[1ex]
{\normalsize \it Institut f\"ur Theoretische Physik, Universit\"at
W\"urzburg, Am Hubland, D-97074~W\"urzburg, Germany}
%}
\vskip 1em
%\vskip 2em
\end{center} \par
\vskip .8cm
\vfil
{\bf Abstract} \par
We study the process $e^{-}e^{+}\to\tilde{\chi}^0_1\tilde{\chi}^0_2$ 
 with the subsequent
decay $\tilde{\chi}^0_2 \to \tilde{\chi}^0_1 \ell^{+}\ell^{-}$
 taking into account the complete spin correlations. 
We give the analytical formulae for the
differential cross section and present numerical results for the
 lepton angular distribution and the
distribution of the opening angle between the outgoing leptons 
 for the LEP2 energy of $\sqrt{s}=193$ GeV and for
 $\sqrt{s}=500$~GeV. We examine 
three representative mixing scenarios in the MSSM and also 
study the influence of
 the common scalar mass parameter $m_0$ on the shape of the
distributions. For the lepton angular distribution the effect of the 
spin correlations amounts to up to 20\% for lower energies.
 We find that the opening angle distribution is  
 suitable for distinguishing between higgsino-like and gaugino-like
neutralinos.
 The shape of the lepton angular distribution is very
sensitive to the mixing in the gaugino sector and to the value
of $m_0$. For higher energies it is also  
suitable for distinguishing between higgsino-like and
gaugino-like neutralino.
\par
\vskip 1cm
\begin{center}
PACS numbers~:~12.60.Jv, 14.80.Ly, 13.88.+e, 13.10.+q, 13.30.Ce
\end{center}
\null
\setcounter{page}{0}
\clearpage
\section{Introduction}
Although the Standard Model (SM) is extraordinarily successful
for describing the electroweak phenomena, it has
 several theoretical shortcomings. The most severe of these deficiencies,
the hierarchy problem, can satisfactorily be solved by the concept of
supersymmetry (SUSY) broken at the TeV scale. This concept received a fresh
impetus by the result that in a SUSY-GUT model the measured coupling
constants evolved to high energies meet at a single point.

The most economical candidate for a realistic SUSY model with minimal gauge
 group $SU(3)\times SU(2)\times U(1)$ and with minimal content of particles
is the Minimal Supersymmetric Extension of the Standard Model (MSSM).
In this paper we consider its simplest version with the conserved 
quantum number R parity.
 Its implications include that SUSY particles can only be produced
in pairs and the lightest supersymmetric particle (LSP) is stable and
escapes detection.
 As usual we assume that this particle is the lightest neutralino
$\tilde{\chi}^0_1$.

Among the new particles the charginos (the supersymmetric partners of the
charged gauge and Higgs bosons) and the neutralinos (the partners of the
neutral gauge and Higgs bosons) are of particular interest.
As they are expected to be lighter than the gluino and in most scenarios
 lighter than the squarks and sleptons \cite{Peskin},\cite{Feng},
the next-to-lightest neutralino
$\tilde{\chi}^0_2$ and the lightest chargino could
be first observed in future experiments at $e^{+} e^{-}$-colliders. In
particular the production of $\tilde{\chi}^0_1 \tilde{\chi}^0_2$ pairs
allows to study a wide region of the SUSY parameter space. Although in
 general chargino production is favoured by larger cross sections, in certain
regions of the parameter space sizeable cross sections for the neutralino
 process can be expected \cite{Bartl2}.
Moreover it might be possible to discover
SUSY by neutralino production if charginos are not accessible.

If new particles are discovered which are possible neutralino candidates,
for a clear identification the complete investigation of their decay
characteristics is indispensable.
Neutralino decay widths and branching ratios have been thoroughly
studied \cite{Mele}. Angular
distributions and angular correlations of the decay products can give
valuable additional information on their composition from photino,
zino and higgsino components.

Another interesting question is to see if angular distributions of decay
products allow separation of neutralino production from chargino production.

Moreover, from decay angular
distributions  with complete spin correlations of the decaying particle
one can infer the spin of the new particles.

Finally the identification of neutralinos would be completed
 by ascertaining their Majorana character. In Ref.
  \cite{Petkov,Bilenky86} 
it is demonstrated that this is possible by means of the
energy distributions of the decay leptons if the neutralinos are produced in
collisions of polarized $e^{+} e^{-}$ beams. The angular distributions of the
decay products might, however, offer the possibility to prove the Majorana
 character if polarized beams are not available. Furthermore the angular
 distributions of the final leptons are suitable observables for studying
 CP-violation in supersymmetric models \cite{Bilenky86},\cite{Oshimo}.

The above mentioned reasons motivate
a study of angular distributions in associated
production of neutralinos and the
subsequent decay of the next-to-lightest neutralino.
 Since angular distributions depend on the polarization of the
parent particles one has to take into account
spin correlations between production and decay. 

In general quantum mechanical interference effects between various 
polarization states of the parent particles
preclude simple factorization of the differential cross section
into a production factor and a decay factor \cite{Kawasaki}
unless the production amplitude is
dominated by a single spin component \cite{Feng}.
However, in Ref. \cite{Tata}
it is shown that the factorization property holds for particles with
spin provided that suitable Lorentz invariant variables are used. But this is
 not the case for angular distributions of the neutralino (or chargino)
decay products in the laboratory frame. For energy distributions
of the final particles in the laboratory frame the spin correlations
are nevertheless usually ignored. 

Heavy fermion production with subsequent decay was 
considered in Ref.~\cite{Weiler}.  
Recently a Monte-Carlo generator for chargino production and decay
including spin correlations was developed in Ref.~\cite{Dionisi}.

In this paper we calculate the cross section of the process
 $e^{+} e^{-}\to\tilde{\chi}^0_1 \tilde{\chi}^0_2$
 and the subsequent direct leptonic decay, $\tilde{\chi}^0_2 \to 
\tilde{\chi}^0_1 \ell^{+} \ell^{-}$. We give the analytical formulae of the 
differential cross section with complete spin correlations of the decaying 
neutralino. We study numerically the influence of these spin correlations 
on energy and angular distributions.   
The energy distributions are independent of spin correlations. However, for 
the lepton angular distributions for lower energies  
the effect of the spin 
correlations amounts to up to 20\%. The shape of the lepton angular
 distribution is very sensitive to the mixing character in the gaugino
 sector and to the
 value of $m_0$. For higher energies it is suitable for
 distinguishing between higgsino- and gaugino-like neutralino. 
 We also analyze the opening angle distributions and 
show that at lower energies they  
are suitable for distinguishing between
higgsino- and gaugino-like neutralinos. This will allow one 
 to constrain the parameter space of the MSSM.
We also consider the influence of the scalar mass parameter $m_0$ 
on the shape of distributions.

In Sec.~2 the general formalism is shown. In Sec.~2.1 the Lagrangian, 
couplings and Feynman diagrams are given, 
 in Sec.~2.2 the spin-density formalism is presented, and in Sec.~2.3
 the kinematics is given. 
In Sec.~3 we present the formulae for the differential
cross section with complete spin correlations. Numerical results for
 the LEP2 energy $\sqrt{s}=193$ GeV and for $\sqrt{s}=500$~GeV and a
discussion are presented in Sec.~4.
\vspace{-.3cm}
\section{General formalism}
\vspace{-.2cm}
\subsection{The Feynman diagrams}
In this section we show the Feynman diagrams and give the Lagrangian for the
 production process,
$e^{-}(k_1) e^{+}(k_2) \to \tilde{\chi}^0_1(q_1) \tilde{\chi}^0_2(q_2)$,
 and for the direct leptonic decay, $\tilde{\chi}^0_2(q_2) \to
\tilde{\chi}^0_1(p_1) \ell^{+}(p_2) \ell^{-}(p_3)$. 
The arguments $k_1, k_2, q_1, q_2$ and $p_1, p_2, p_3$ denote
the momenta of the
incoming electron, positron, the produced neutralinos
$\tilde{\chi}^0_1,\tilde{\chi}^0_2$ and the outgoing neutralino
$\tilde{\chi}^0_1$ and leptons $\ell^{+}, \ell^{-}$
from the $\tilde{\chi}^0_2$ decay.
Both the production and
the decay process contain contributions from $Z^0$ exchange in the
direct channel ($s$-channel) and from $\tilde{e}_L$ and $\tilde{e}_R$
 exchange in the crossed channels ($t$-, $u$-channel). 
 We introduce the kinematic variables:
\begin{eqnarray}
s & = & (k_1+k_2)^2, \quad\quad\quad \bar{s}=(p_2+p_3)^2,\\
t & = & (k_2-q_2)^2, \quad\quad\quad \bar{t}=(q_2-p_2)^2,\\ \nopagebreak
u & = & (k_1-q_2)^2, \quad\quad\quad \bar{u}=(q_2-p_3)^2.
\end{eqnarray}
Channels referring to the decay are marked by a dash. The Feynman diagrams 
are shown in FIG.\ref{fig_1}.

From the interaction Lagrangian of the MSSM ( in our notation and
conventions we follow closely \cite{Haber-Kane}),
\begin{eqnarray}
& & {\cal L}_{Z^0 \ell^{+} \ell^{-}} =
-\frac{g}{\cos\theta_W}Z_{\mu}\bar{\ell}\gamma^{\mu}[L_{\ell}P_L+
 R_{\ell}P_R]\ell+\mbox{h.c.} \\
& & {\cal L}_{Z^0\tilde{\chi}^0_i\tilde{\chi}^0_j} =
\frac{1}{2}\frac{g}{\cos\theta_W}Z_{\mu}\bar{\tilde{\chi}}^0_i\gamma^{\mu}
[O_{ij}^L P_L+O_{ij}^R P_R]\tilde{\chi}^0_j+\mbox{h.c.}\\
& & {\cal L}_{\ell \tilde{\ell}\tilde{\chi}^0_i} =
g f^L_{\ell i}\bar{\ell}P_R\tilde{\chi}^0_i\tilde{\ell}_L+
g f^R_{\ell i}\bar{\ell}P_L\tilde{\chi}^0_i\tilde{\ell}_R+\mbox{h.c.},
 \quad i, j=1,\cdots,4,
\end{eqnarray}
%\newpage

we obtain the couplings:
\begin{itemize}
\item $L_{\ell}=T_{3\ell}-e_{\ell}\sin^2\theta_W, \quad
 R_{\ell}=-e_{\ell}\sin^2\theta_W$
\item
$f_{\ell i}^L = -\sqrt{2}\bigg[\frac{1}{\cos
\theta_W}(T_{3\ell}-e_{\ell}\sin^2\theta_W)N_{i2}+e_{\ell}\sin \theta_W
N_{i1}\bigg]$,\\
$f_{\ell i}^R = -\sqrt{2}e_{\ell} \sin \theta_W\Big[\tan \theta_W N_{i2}^*-
                                                           N_{i1}^*\Big]$
\item
$O_{ij}^L=-\frac{1}{2}\Big(N_{i3}N_{j3}^*-N_{i4}N_{j4}^*\Big)\cos2\beta
  -\frac{1}{2}\Big(N_{i3}N_{j4}^*+N_{i4}N_{j3}^*\Big)\sin2\beta$,\\
$O_{ij}^R=-O_{ij}^{L*}$.
\end{itemize}
In our case $i=1$, $j=2$ and we shall write $O^{L,R}_{12}=O^{L,R}$.
Here $P_{L, R}=\frac{1}{2}(1\mp \gamma_5)$, $g$ is the weak coupling
constant ($g=e/\sin\theta_W, e>0$), and $e_\ell$ and $T_{3 \ell}$ denote the
charge and the third component of the weak isospin of the lepton
$\ell$. Furthermore $\tan \beta=\frac{v_2}{v_1}$,
 $v_{1,2}$ are the vacuum expectation values of the two neutral Higgs fields
and $N_{ij}$ is the unitary $4\times 4$ matrix which diagonalizes
the neutral gaugino-higgsino mass matrix in the basis $\tilde{\gamma},
\tilde{Z}, \tilde{H}^0_a, \tilde{H}^0_b$.
Since we disregard in this paper CP violating phenomena the elements
$N_{ij}$ of the diagonalized matrix and the couplings
 can be chosen real.
Then some of the neutralino
mass eigenvalues may be negative and we shall write them in the form
$\eta_i m_i$ with $m_i>0, \eta_i=\pm 1$ \cite{fraas1}.
The respective amplitudes for the Feynman diagrams are taken from 
Ref.~\cite{Fraas2}.
\subsection{Spin density matrix for production and decay}
The differential cross section for the production  
$e^{-}(k_1)e^{+}(k_2) \to \tilde{\chi}^0_1(q_1) \tilde{\chi}^0_2(q_2)$ and
the subsequent decay $\tilde{\chi}^0_2(q_2) \to \tilde{\chi}^0_1(p_1) 
\ell^{+}(p_2) \ell^{-}(p_3)$ 
is given by
\begin{equation}
d\sigma = \frac{1}{2 \sqrt{\lambda(s, m_e^2, m_e^2)}}
|T^{\lambda_{\bar{1}} \lambda_{1} \lambda_{+} \lambda_-}_{\lambda_{e^{-}} 
\lambda_{e^{+}}}|^2
(2\pi)^4 \delta^{(4)}(k_1+k_2-q_1-p_1-p_2-p_3) dlips(q_1,p_1,p_2,p_3),
\end{equation}
where $s$ is the cms-energy squared of the incoming $e^{-}$ and $e^{+}$ and
$\lambda(x, y, z)=x^2+y^2+z^2-2xy-2xz-2yz$ is
the kinematical triangle function. All lepton masses
 are neglected.
In Eq.~(7) 
$\lambda_{e^{-}}, \lambda_{e^{+}}, \lambda_{\bar{1}}$ and
 $\lambda_1, \lambda_{+}, \lambda_{-}$ 
 label the helicities of the electron, positron and $\tilde{\chi}^0_1$
 of the production process and the helicities
 of the $\tilde{\chi}^0_1$, $\ell^{+}, \ell^{-}$ of the decay process, 
respectively.
$dlips(q_1,p_1,p_2,p_3)=\frac{d^3 q_1}{(2  \pi)^3 2 q^0_1}\cdots
\frac{d^3 p_3}{(2 \pi)^3 2 p^0_3}$
is the Lorentz invariant phase space element of the four final particles.
 The helicities $\lambda_{e^{-}}, \lambda_{e^{+}}$ 
of the initial particles are 
averaged because in this paper we consider the case of 
unpolarized beams.
Since the polarization of the outgoing particles will not be measured, the
 helicities of the final particles $\lambda_{\bar{1}}$ and 
$\lambda_1, \lambda_{+}, \lambda_{-}$ have to be summed over. 
Therefore we suppress these helicity indices in the following and 
 only display the helicity $\lambda_2$ of $\tilde{\chi}^0_2$.

The total width $\Gamma_2$ of the decaying neutralino
$\tilde{\chi}^0_2$ is small
 compared to its mass $m_2$. Therefore  
 the amplitude $T$ of the combined process is
a coherent sum over all polarization states of the helicity amplitude
$P^{\lambda_2}$ for the production process times the helicity amplitude
 $D_{\lambda_2}$ for the decay process and a pseudopropagator
$\Delta_2=\frac{1}{q_2^2-m_2^2+i m_2 \Gamma_2}$ of $\tilde{\chi}^0_2$:
\begin{equation}
T=\Delta_2 P^{\lambda_2} D_{\lambda_2}
\end{equation}
The amplitude squared for the combined process
\begin{equation}
|T|^2=|\Delta_2|^2 \rho^{\lambda_2 \lambda_2'}_P
\rho^D_{\lambda_2' \lambda_2} \label{N}
\end{equation}
 is thus composed from the
unnormalized spin density matrix
\begin{equation}
\rho^{\lambda_2 \lambda_2'}_P=P^{\lambda_2} P^{\lambda_2' *}
\end{equation}
of the neutralino $\tilde{\chi}^0_2$ and
 the decay matrix
\begin{equation}
\rho^{D}_{\lambda_2' \lambda_2}=D_{\lambda_2} D_{\lambda_2'}^{*}
\end{equation}
for the respective decay channel.
 As in FIG.\ref{fig_2} all helicity indices 
but that of the decaying neutralino are
 suppressed. Repeated indices are summed over.

Squaring the total amplitude one obtains interference terms between
various helicity amplitudes. These terms preclude factorization in a production
factor  $\sum_{\lambda_2} |P^{\lambda_2}|^2$ times a decay factor
$\overline{\sum}_{\lambda_2} |D_{\lambda_2}|^2$ as for the case of spinless
particles.

We use the general formalism to calculate
 the helicity amplitudes for production
 and decay of a particle
with four-momentum $p$ and mass $m$.
Therefore we introduce three spacelike four-vectors
$s^a_{\mu},  (a=1,2,3)$ which together with $p/m$ form an orthonormal set:
\begin{eqnarray}
\frac{p}{m} \cdot s^{a} & = & 0 \\
s^{a} \cdot s^{b} & = & -{\delta}^{ab} \\
s^{a}_{\mu}\cdot s^{a}_{\nu} & = & -g_{\mu\nu}+\frac{p_{\mu}p_{\nu}}{m^2}.
\label{A}\end{eqnarray}
A convenient choice for the explicit form of $s^a$ is
 in a coordinate system where the direction of the three-momentum
is $\hat{p}=(\sin{\Theta}\cos{\Phi},\sin{\Theta}\sin{\Phi},
\cos{\Theta})$ \cite{Haber}:
\begin{eqnarray}
s^{1\mu} & = & (0,-\sin{\Phi},\cos{\Phi},0)\\
s^{2\mu} & = &
(0,\cos{\Theta}\cos{\Phi},\cos{\Theta}\sin{\Phi},-\sin{\Theta})\\
s^{3\mu} & = & \frac{1}{m}(\left|\vec{p}\right|,E \hat{p}).
\end{eqnarray}
Then in this frame of reference $s^{(1),(2)}$ and
$s^{(3)}$ describe transverse and longitudinal pola\-rization of the particle.

When computing the density matrices for production and
decay, Eqs.~(10),~(11), we make
 use of the Bouchiat-Michel formulae \cite{Haber}:
\begin{equation}
u(p,\lambda')\bar{u}(p,\lambda)=\frac{1}{2}
 [\delta_{\lambda{\lambda '}}+\gamma_5 \not\!s^{a}
  \sigma^{a}_{\lambda{\lambda '}}](\not\!p +m) \label{3L1}
\end{equation}
\vspace*{-.3cm}
\begin{equation}
v(p,\lambda')\bar{v}(p,\lambda)=\frac{1}{2}
 [\delta_{\lambda{\lambda '}}+\gamma_5 \not\!s^{a}
 \sigma^{a}_{{\lambda '}\lambda}](\not\!p-m). \label{3L2}
\end{equation}
In the amplitude squared (Eq.~(9)) the spin vectors $s^{a}$ 
linearly enter the  matrices
$\rho_P$ (Eq.~(10)) and $\rho^D$ (Eq.~(11)). They induce by Eq.~(14)
the above mentioned quantum mechanical
correlations between production and decay.

\subsection{Kinematics and Phase Space}
We split the phase space into the one for the production, 
$e^{-}(k_1) e^{+}(k_2) \to \tilde{\chi^0_1}(q_1) \tilde{\chi}^0_2(q_2)$, 
and the one for the three
particle decay, $\tilde{\chi}^0_2(q_2) \to \tilde{\chi}^0_1(p_1) \ell^{+}(p_2) 
\ell^{-}(p_3)$.
 Then we obtain
the differential cross section in the $e^{-}e^{+}$--cms by integrating
over the effective mass squared $s_2$ of the decaying neutralino 
$\tilde{\chi}^0_2$ \cite{Pilkuhn}:
\begin{eqnarray}
d\sigma & = & \frac{1}{8E_b^2}\int \frac{ds_2}{2\pi}
\frac{W}{(s_2-m_2^2)^2+\Gamma_2^2 m_2^2}\nonumber\\
& \times &  (2 \pi)^4 \delta^{(4)}(q_1+q_2-k_1-k_2)
 \frac{d^3q_1}{(2\pi)^3 2E_1} \frac{d^3q_2}{(2\pi)^3 2E_2}\nonumber\\
& \times & (2\pi)^4 \delta^{(4)}(p_1+p_2+p_3-q_2)
\frac{d^3p_1}{ (2\pi)^3 2\bar{E}_1}
\frac{d^3p_2}{ (2\pi)^3 2 E_{+}}
\frac{d^3p_3}{ (2\pi)^3 2 E_{-}}
\end{eqnarray}
with $W= \rho_{P}^{\lambda_2 \lambda'_2} \rho^D_{\lambda'_{2}\lambda_2}$
(compare Eq.~(9) ). The energies of the produced $\tilde{\chi}^0_1,
 \tilde{\chi}^0_2$ are
$E_1, E_2$ and those of $\tilde{\chi}^0_1, \ell^{+}$ and $\ell^{-}$ from
 the decay
 are denoted by $\bar{E}_1, E_{+}$ and $E_{-}$. $E_b$ is the beam
energy, $m_2$ is the mass of $\tilde{\chi}^0_2$ and the mass
 of $\tilde{\chi}^0_1$
is labeled by $m_1$. Finally $\Theta_{M-}$ ($\Theta_{M+}$) is the
angle between the electron beam and the outgoing negatively (positively)
 charged lepton $\ell^{-}$ $(\ell^{+})$ in the laboratory frame 
(FIG.\ref{fig_3}).

If the total width of the decaying particle $\tilde{\chi}^0_2$ 
is much smaller than its mass,
 $\Gamma_2\ll m_2$, one can make the narrow width approximation
$ \frac{1}{(s_2-m_2^2)^2+m_2^2 \Gamma_2^2}
\approx \frac{\pi}{m_2 \Gamma_2} \delta(s_2-m_2^2)$
and one obtains for
the differential cross section in the $e^{+}e^{-}$--cms, i.e. in the
laboratory frame:
\begin{eqnarray}
d\sigma &=& W
\frac{1}{16 (2\pi)^7}\frac{1}{|m_2|\Gamma_2}
\frac{|\vec{q}_2|}{64 E_b^3}
\frac{E_{+} E_{-} dE_{-} \sin\Theta d\Theta d\Phi d\Omega_{+}
 d\Omega_{-}}
     {[E_2-|\vec{q}_2|\cos\theta_{2+}-E_{-}(1-\cos\Theta_{+-})]}.
\end{eqnarray}
$\Theta$ is the production angle of $\tilde{\chi}^0_2$, $\Phi$ describes 
the rotation around the electron beam axis, and  
$d\Omega_+ (d\Omega_-)$ is the solid angle of $\ell_+ (\ell_-)$.

In consequence of momentum conservation the energy $E_{+}$ 
 is determined by $E_{-}$, the angles
 $\theta_{2-} (\theta_{2+})$ between $\ell^{-}$ ($\ell^{+}$) and
$\tilde{\chi}^0_2$,
and the opening angle $\Theta_{+-}$ between the leptons:
\begin{equation}
E_{+}=
 \frac{m_2^2-m_1^2-2 E_{-}(E_2-|\vec{q}_2|\cos\theta_{2-})}
 {2[E_2-|\vec{q}_2|\cos\theta_{2+}-E_{-}(1-\cos\Theta_{+-})]}.
\end{equation}

The possible region of the lepton energy $E_{-}$ depends on the lepton
decay angle $\theta_{2-}$:
\begin{equation}
0\le E_{-} \le \frac{m_2^2-m_1^2}{2 (E_2-|\vec{q_2}| \cos\theta_{2-})},
\end{equation}
so that $E_{+}=0$, see Eq.~(22), at the upper bound.

\section{Analytical formulae}
In this section we give the analytical expressions for the product
$W=\rho_P^{\lambda_2 \lambda'_2} \rho^D_{\lambda'_2 \lambda_2}$ 
(compare Eq.~(9)) of the density matrices for the production process, 
$e^{-}(k_1) e^{+}(k_2) \to \tilde{\chi}^0_1(q_1) \tilde{\chi}^0_2(q_2)$, 
 and the direct leptonic decay, $\tilde{\chi}^0_2(q_2) \to
\tilde{\chi}^0_1(p_1) \ell^{+}(p_2) \ell^{-}(p_3)$. Both the production and
the decay process contain contributions from $Z^0$ exchange in the
direct channel and from $\tilde{e}_L$ and $\tilde{e}_R$
 exchange in the crossed channels (compare FIG.\ref{fig_1}). 

The product $W$ in Eq.~(9) is a sum of
contributions from the different production and decay amplitudes and
their interference terms,  
\begin{equation}
W=\sum_{ab, cd} W_{ab, cd},
\end{equation}
where the first pair $a, b$ of indices refers to the production
process, and the second pair $c, d$ refers to the decay process.
 The pairs (ab) and (cd) run through all
combinations of the values $Z, L_t, L_u, R_t, R_u$, 
where $Z$ and $L, R$ denote the exchanged particles $Z^0$ and $\tilde{e}_L,
 \tilde{e}_R$ and $\tilde{\ell}_L, \tilde{\ell}_R$, respectively. 
 In the case of slepton exchange the indices $t, u$ denote the channel.
Thus $W_{Z L_t,R_t R_u}$ is the
contribution of the interference term to the production process from
 $Z^0$-exchange and
 $\tilde{e}_L$-exchange in the t-channel 
  and the interference term to the decay process
from $\tilde{\ell}_R$-exchange in the $\bar{t}$-channel and
$\tilde{\ell}_R$-exchange in the $\bar{u}$-channel (FIG.\ref{fig_1}).

There are altogether 121 contributions $W_{ab, cd}$ which can be classified
into 16 groups. We give one representative of each group and list the 
indices of the possible combinations.
 The other terms $W_{ab, cd}$ of this group are then
obtained by substituting the propagator combination according to the pair of
 indices (see Eqs. 
(65)--(70)) and by substituting momenta and couplings and sign factors 
as given in TABLE~I below.
The substitutions of the momenta $k_1, k_2, p_2, p_3$ 
and of the couplings $O^{L}, O^{R}, L, R$ 
also have to be performed in Eqs. (41)--(64).
In the representative term the sign factors $\mu,
\nu, \omega, \tau, \upsilon$ have the value +1.

For illustration we consider as an example 
  $ W_{L_tL_u,L_tL_t}$,
 which is the representative of Group~12. 
In order to get the contribution of $W_{R_tR_u,L_uL_u}$ 
one has to change the propagator combination
$\Delta_L(t)\Delta_L(u)\Delta_L(\bar{t})\Delta_L(\bar{t})\to
\Delta_R(t)\Delta_R(u)\Delta_L(\bar{u})\Delta_L(\bar{u})$ and one has to
 substitute $\tilde{e}_L \to \tilde{e}_R$ in the production process  
 and one has to substitute 
 $\bar{t}\to\bar{u}$ in the decay process. When substituting 
$\tilde{e}_L\to\tilde{e}_R$, one has to change the sign factors 
$\nu,\omega,\upsilon$ and the couplings according to the first line 
of TABLE~I. Moreover, when substituting $\bar{t}\to\bar{u}$ one has to change
 the sign factors $\tau,\upsilon$, the couplings and the 
momenta $p_2\to p_3$ according to the fourth line of TABLE~I.

\vspace{.5cm}
{\it Group 1 (1 term):}\\
\vspace{-.5cm}
\begin{eqnarray}
W_{ZZ,ZZ} & = & 32 |\Delta_Z(s)|^2 |\Delta_Z(\bar{s})|^2
\Big\{D1\cdot P1-D2_3\cdot S1_2+D2_2\cdot S1_3\Big\}
\end{eqnarray}

\vspace{.5cm}
{\it Group 2 (4 terms):}\\
Production: $(L_tZ),(R_tZ),(L_uZ),(R_uZ)$ 
\hspace{1cm}Decay: $(ZZ)$\\
\vspace{-.5cm}
\begin{eqnarray}
W_{L_tZ,ZZ} & = & 16 \Delta_L(t) |\Delta_Z(\bar{s})|^2
\Big\{ - \mu Re\big(\Delta_Z(s)\big) [D1\cdot P2 + \upsilon (D2_3 \cdot S2_2
                                                 - D2_2 \cdot S2_3)]\nonumber\\
& &\mbox{\hspace{2.6cm}}
       + \nu \upsilon Im\big(\Delta_Z(s)\big)
 [D2_3\cdot S6_2-D2_2\cdot S6_3]\Big\}
\end{eqnarray}

\vspace{.5cm}
{\it Group 3 (4 terms):}\\
Production: $(L_tL_t),(R_tR_t),(L_uL_u),(R_uR_u)$ \quad Decay: $(ZZ)$\\
\vspace{-.5cm}
\begin{eqnarray}
W_{L_tL_t,ZZ}&  = &
 8 \Delta_L^2(t) |\Delta_Z(\bar{s})|^2
\Big\{ (k_1q_1)(k_2q_2) D1-
       \upsilon (D2_3\cdot S3_2- D2_2\cdot S3_3)\Big\}
\end{eqnarray}

\vspace{.5cm}
{\it Group 4 (2 terms):}\\
Production: $(L_tL_u),(R_tR_u)$ \hspace{3.3cm}Decay: $(ZZ)$\\
\vspace{-.5cm}
\begin{eqnarray}
W_{L_tL_u,ZZ} & = &  -8 \Delta_L(t)\Delta_L(u)|\Delta_Z(\bar{s})|^2
\Big\{ D1\cdot P4 + \upsilon \eta_1 m_1 (D2_3\cdot S4_2-D2_2\cdot S4_3)
\Big\}
\end{eqnarray}

\vspace{.5cm}
{\it Group 5 (4 terms):}\\
Production: $(ZZ)$ \hspace{5cm}Decay: $(L_tZ),(R_tZ),(L_uZ),(R_uZ)$\\
\vspace{-.5cm}
\begin{eqnarray}
W_{ZZ,L_tZ} & = &  16 |\Delta_Z(s)|^2 \Delta_L(\bar{t})
\tau \Big\{ Re(\Delta_z(\bar{s})) \{P1\cdot D3 -\upsilon
[D4\cdot S1_2-D5\cdot S1_3]\}\nonumber\\
& &\mbox{\hspace{2.8cm}}
-Im(\Delta_Z(\bar{s})) (R_e^2-L_e^2)O^L \eta_1 m_1
[S7_2\cdot P3_1-S7_1\cdot P3_2]\Big\}
\end{eqnarray}

\vspace{.5cm}
{\it Group 6 (16 terms):}\\
Production: $(L_tZ),(R_tZ),(L_uZ),(R_uZ)$ 
\hspace{.9cm} Decay: $(L_tZ),(R_tZ),(L_uZ),(R_uZ)$\\
\vspace{-.5cm}
\begin{eqnarray}
W_{L_tZ,L_tZ} & = & 8\Delta_L(t)\Delta_L(\bar{t}) \tau \nonumber\\
& \cdot &
\Big\{-\mu Re(\Delta_Z(s)) Re(\Delta_Z(\bar{s}))
 [P2\cdot D3+\upsilon( D4\cdot S2_2 -D5\cdot S2_3)]
\nonumber\\
& &
 +\omega \upsilon Im(\Delta_Z(s)) Re(\Delta_Z(\bar{s})) 
[D4\cdot S6_2-D5\cdot S6_3]\nonumber\\
& &
 -\nu Re(\Delta_Z(s)) Im(\Delta_Z(\bar{s})) L_e \eta_1
m_1\nonumber\\
& &\mbox{\hspace{.2cm}}
\cdot[2\eta_2 m_2O^R(k_1q_1)
S7_2+\eta_1 m_1 O^{L}(k_1q_2)
S7_2 - \eta_1 m_1 O^{L} (k_2q_2) S7_1]\nonumber\\
& &
 + Im(\Delta_Z(s)) Im(\Delta_Z(\bar{s}))L_e O^L \bar{L}_e\bar{O}^R m_1^2
 [(p_2q_2) S4_3-(p_3q_2) S4_2 +m_2^2 S5]\Big\}
\end{eqnarray}

\vspace{.5cm}
{\it Group 7 (16 terms)}:\\
Production: $(L_tL_t),(R_tR_t),(L_uL_u),(R_uR_u)$ 
\hspace{.5cm} Decay: $(L_tZ),(R_tZ),(L_uZ),(R_uZ)$\\
\vspace{-.5cm}
\begin{eqnarray}
W_{L_tL_t,L_tZ} & = & 4\Delta^2_L(t)\Delta_L(\bar{t})
\tau \Big\{Re(\Delta_Z(\bar{s}))[(k_1q_1)(k_2q_2) D3-
\upsilon(D4\cdot S3_2-D5\cdot S3_3)]
\nonumber\\
& &\mbox{\hspace{2.2cm}}
+\omega Im(\Delta_Z(\bar{s})) \eta_1 m_1 \eta_2 m_2(k_1q_1)S7_2\Big\}
\end{eqnarray}

{\it Group 8 (8 terms)}:\\
Production: $(L_tL_u),(R_tR_u)$ 
\hspace{3.4cm} Decay: $(L_tZ),(R_tZ),(L_uZ),(R_uZ)$\\
\vspace{-.5cm}
\begin{eqnarray}
W_{L_tL_u,L_tZ} & = &-4\Delta_L(t)\Delta_L(u) \Delta_L(\bar{t})
%\nonumber\\& & 
\cdot\tau
\Big\{ Re(\Delta_Z(\bar{s}))[D3\cdot P4
+\upsilon \eta_1 m_1(D4\cdot S4_2-D5\cdot S4_3)]
\nonumber\\
& &\mbox{\hspace{3.7cm}}
      + \omega
 Im(\Delta_Z(\bar{s})) m_1^2[(k_1q_2) S7_2-(k_2q_2) S7_1] \Big\}
\end{eqnarray}

\vspace{.5cm}
{\it Group 9 (4 terms):}\\
Production: $(ZZ)$ 
\hspace{5.2cm} Decay: $(L_tL_t),(R_tR_t),(L_uL_u),(R_uR_u)$\\
\vspace{-.5cm}
\begin{eqnarray}
W_{ZZ,L_tL_t} & = & 8 |\Delta_Z(s)|^2 \Delta_L^2(\bar{t})(p_1p_3)
\cdot \Big\{(p_2q_2) P1+\upsilon \eta_2 m_2 S1_2\Big\}
\end{eqnarray}

\vspace{.5cm}
{\it Group 10 (16 terms):}\\
Production: $(L_tZ),(R_tZ),(L_uZ),(R_uZ)$ 
\hspace{1cm}Decay: $(L_tL_t),(R_tR_t),(L_uL_u),(R_uR_u)$\\
\vspace{-.5cm}
\begin{eqnarray}
W_{L_tZ,L_tL_t} & = & 4 \Delta_L(t)\Delta_L^2(\bar{t})(p_1p_3)
%\nonumber\\& & 
\cdot
\Big\{ - \mu Re(\Delta_Z(s)) [(p_2q_2) P2-\upsilon \eta_2 m_2 S2_2]
       - \nu \upsilon Im(\Delta_Z(s)) \eta_2 m_2 S6_2\Big\}
\end{eqnarray}

\vspace{.5cm}
{\it Group 11 (16 terms):}\\
Production: $(L_tL_t),(R_tR_t),(L_uL_u),(R_uR_u)$ 
\quad Decay: $(L_tL_t),(R_tR_t),(L_uL_u),(R_uR_u)$\\
\vspace{-.5cm}
\begin{eqnarray}
W_{L_tL_t,L_tL_t} & = & 2 \Delta_L^2(t)\Delta_L^2(\bar{t})
\Big\{ (p_1p_3)(p_2q_2) (k_1q_1)(k_2q_2)+\upsilon
                                              \eta_2 m_2 (p_1p_3) S3_2
\Big\}
\end{eqnarray}

\vspace{.5cm}
{\it Group 12 (8 terms):}\\
Production: $(L_tL_u),(R_tR_u)$ 
\hspace{3.2cm} Decay: $(L_tL_t),(R_tR_t),(L_uL_u),(R_uR_u)$\\
\vspace{-.5cm}
\begin{eqnarray}
W_{L_tL_u,L_tL_t} & = & -2 \Delta_L(t)\Delta_L(u)\Delta_L^2(\bar{t})
\Big\{ (p_1p_3)(p_2q_2) P4-
         \upsilon \eta_1 m_1\eta_2 m_2 (p_1p_3) S4_2\Big\}
\end{eqnarray}

\vspace{.5cm}
{\it Group 13 (2 terms):}\\
Production: $(ZZ)$ 
\hspace{5cm}Decay: $(L_tL_u),(R_tR_u)$\\
\vspace{-.5cm}
\begin{eqnarray}
W_{ZZ,L_tL_u} & = & 8 |\Delta_Z(s)|^2\Delta_L(\bar{t})\Delta_L(\bar{u})
\Big\{\eta_1 m_1 \eta_2 m_2 (p_2 p_3) P1
%\nonumber\\
%& & \mbox{\hspace{3.7cm}}
+\upsilon \eta_1 m_1 [(p_3q_2) S1_2-(p_2q_2)  S1_3]\Big\}
\end{eqnarray}

\vspace{.5cm}
{\it Group 14 (8 terms):}\\
Production: $(L_tZ),(R_tZ),(L_uZ),(R_uZ)$  
\hspace{1.1cm}Decay: $(L_tL_u),(R_tR_u)$\\
\vspace{-.5cm}
\begin{eqnarray}
W_{L_tZ,L_tL_u} & = & 4\Delta_L(t)\Delta_L(\bar{t})\Delta_L(\bar{u})
%\nonumber\\& &
\cdot\Big\{- \mu Re(\Delta_Z(s))[\eta_1 m_1 \eta_2 m_2 (p_2p_3) P2
                 +\upsilon \eta_1 m_1(-(p_3q_2) S2_2+(p_2q_2) S2_3)]
\nonumber\\
& & \mbox{\hspace{3.3cm}}
+ \nu \upsilon Im(\Delta_Z(s)) \eta_1 m_1[-(p_3q_2) S6_2+(p_2q_2) S6_3]\Big\}
\end{eqnarray}

\vspace{.5cm}
{\it Group 15 (8 terms):}\\
Production: $(L_tL_t),(R_tR_t),(L_uL_u),(R_uR_u)$  
\hspace{.5cm}Decay: $(L_tL_u),(R_tR_u)$\\
\vspace{-.5cm}
\begin{eqnarray}
W_{L_tL_t,L_tL_u} & = & 2 \Delta_L^2(t) \Delta_L(\bar{t})\Delta_L(\bar{u})
\eta_1 m_1 \eta_2 m_2 (k_1q_1)
%\nonumber\\& &
\cdot \Big\{ (p_2p_3)(k_2q_2)+\upsilon[-(p_3q_2)(k_2p_2)+(p_2q_2)(k_2p_3)]
\Big\}
\end{eqnarray}

\vspace{.5cm}
{\it Group 16 (4 terms):}\\
Production: $(L_tL_u),(R_tR_u)$  
\hspace{3.4cm}Decay: $(L_tL_u),(R_tR_u)$\\
\vspace{-.5cm}
\begin{eqnarray}
W_{L_tL_u,L_tL_u} & = & -2 \Delta_L(t)\Delta_L(u)\Delta_L(\bar{t})
\Delta_L(\bar{u}) \eta_1 m_1
%\nonumber\\& &
\cdot\Big\{ \eta_2 m_2 (p_2p_3)  P4 +\upsilon \eta_1 m_1 [-(p_3q_2) S4_2
                                            +(p_2q_2) S4_3]\Big\}
\end{eqnarray}

For the sake of a clear presentation of our analytical results we
have introduced three groups of abbreviations. The first group refers to the
production process:
\begin{eqnarray}
& & P1 = (R_e^2+L_e^2)\{ O^{L^2}
        [(k_1q_1)(k_2q_2)+(k_1q_2)(k_2q_1)]+O^L O^R
             \eta_1 m_1\eta_2 m_2(k_1k_2)\}\\
& & P2 = 2 L_e O^R (k_1q_1)(k_2q_2) + \eta_1 m_1\eta_2 m_2L_e O^L (k_1k_2) \\
& & P3_1 = \eta_1 m_1 O^R (k_1q_2)+\eta_2 m_2O^L(k_1q_1)\\
& & P3_2 = \eta_1 m_1 O^R (k_2q_2)+\eta_2 m_2O^L(k_2q_1)\\
& & P4 = \eta_1 m_1 \eta_2 m_2 (k_1k_2)
\end{eqnarray}
and the second group refers to the decay process:
\begin{eqnarray}
& & D1 = (\bar{R}_e^2+\bar{L}_e^2)\bar{O}^{L^2}
       \{(p_1p_3)(p_2q_2)+(p_1p_2)(p_3q_2)+\eta_1 m_1\eta_2 m_2(p_2p_3)\}\\
& & D2_2 = (\bar{R}_e^2-\bar{L}_e^2)\bar{O}^{L^2}
         \{\eta_1 m_1(p_2q_2)+\eta_2 m_2(p_1p_2)\}\\
& & D2_3 = (\bar{R}_e^2-\bar{L}_e^2)\bar{O}^{L^2}
         \{\eta_1 m_1(p_3q_2)+\eta_2 m_2(p_1p_3)\}\\
& & D3 = \bar{L}_e[2\bar{O}^L(p_1p_3)(p_2q_2)
                 -\bar{O}^R\eta_1 m_1 \eta_2 m_2(p_2p_3)]\\
& & D4 = \bar{L}_e [\bar{O}^R \eta_1 m_1 (p_3q_2)-2
                  \bar{O}^L \eta_2 m_2 (p_1p_3)]\\
& & D5 = \bar{L}_e\bar{O}^R\eta_1 m_1(p_2q_2).
\end{eqnarray}
The third class is related to the spin correlations between production
and decay and connects the momenta of both subprocesses:
\begin{eqnarray}
& & S1_2 = (R_e^2-L_e^2) \{\eta_1 m_1O^RO^L[(k_2p_2)(k_1q_2)-(k_1p_2)(k_2q_2)]
\nonumber\\
& &
 +\eta_2 m_2O^{L^2}[(k_2q_1)\Big(-(k_1p_2)+\frac{(k_1q_2)(p_2q_2)}{m_2^2}\Big)
           -(k_1q_1)\Big(-(k_2p_2)+\frac{(k_2q_2)(p_2q_2)}{m_2^2}\Big)]\}
\\
& & S1_3 = (R_e^2-L_e^2) \{\eta_1
m_1O^RO^L[(k_2p_3)(k_1q_2)-(k_1p_3)(k_2q_2)]\nonumber\\
& &
+\eta_2 m_2O^{L^2}[(k_2q_1)\Big(-(k_1p_3)+\frac{(k_1q_2)(p_3q_2)}{m_2^2}\Big)
           -(k_1q_1)\Big(-(k_2p_3)+\frac{(k_2q_2)(p_3q_2)}{m_2^2}\Big)\}
\\
& & S2_2 = -2\eta_2 m_2L_eO^R(k_1q_1)[-(k_2p_2)+\frac{(k_2q_2)(p_2q_2)}{m_2^2}]
\nonumber\\
& &\hspace{1.3cm}
      +\eta_1 m_1L_eO^L[(k_2p_2)(k_1q_2)-(k_1p_2)(k_2q_2)]\\
& & S2_3 = -2\eta_2 m_2L_eO^R(k_1q_1)[-(k_2p_3)+\frac{(k_2q_2)(p_3q_2)}{m_2^2}]
\nonumber\\
& &\hspace{1.3cm}
           +\eta_1 m_1L_eO^L[(k_2p_3)(k_1q_2)-(k_1p_3)(k_2q_2)]\\
& & S3_2 = \eta_2 m_2(k_1q_1)[-(k_2p_2)+\frac{(k_2q_2)(p_2q_2)}{m_2^2}]\\
& & S3_3 =\eta_2 m_2(k_1q_1)[-(k_2p_3)+\frac{(k_2q_2)(p_3q_2)}{m_2^2}]\\
& & S4_2 = (k_2p_2)(k_1q_2)-(k_1p_2)(k_2q_2)\\
& & S4_3 = (k_2p_3)(k_1q_2)-(k_1p_3)(k_2q_2)\\
& & S5 = (k_1p_3)(k_2p_2)-(k_1p_2)(k_2p_3)\\
& & S6_2 = L_eO^L\eta_1 m_1[k_2k_1p_2q_2]\\
& & S6_3 = L_eO^L\eta_1 m_1[k_2k_1p_3q_2]\\
& & S7_1 = \bar{L}_e\bar{O}^R[q_2k_1p_3p_2]\\
& & S7_2 = \bar{L}_e\bar{O}^R[q_2k_2p_3p_2]
\end{eqnarray}
with $[abcd]=\epsilon_{\mu \nu \rho \sigma}a^{\mu}b^{\nu}c^{\rho}
d^{\sigma}$.

\vspace{.5cm}
These terms, i.e. Eqs.~(52)--(64), would be missing if we had assumed
factorization of the differential cross section in production and decay.

For a better transparency all couplings originating from the decay
process are marked by a dash.
The indices are used in order to emphasize the symmetry between $e^{-}$ and
$e^{+}$ in the initial state and $\ell^{-}$ and $\ell^{+}$ 
in the final state, respectively.

We have introduced the following products of propagators and coupling
constants:
\begin{eqnarray}
\Delta_Z (s) & = & \frac{g^2}{\cos^2\theta_W}\cdot
       \frac{1}{s-m_Z^2+i m_Z \Gamma_Z}\\
\Delta_L(t) & = & \frac{g^2}{t-m_{\tilde{e}_L}^2}\cdot f_{\ell 1}^{L*}\cdot
                                             f_{\ell 2}^L
\quad\quad,\quad\quad
\Delta_R(t) = \frac{g^2}{t-m_{\tilde{e}_R}^2}\cdot f_{\ell 1}^{R*}\cdot
                                             f_{\ell 2}^R\\
\Delta_L(u) & = & \frac{g^2}{u-m_{\tilde{e}_L}^2}\cdot f_{\ell 1}^L\cdot
                                             f_{\ell 2}^{L*}
\quad\quad,\quad\quad
\Delta_R(u) = \frac{g^2}{u-m_{\tilde{e}_R}^2}\cdot f_{\ell 1}^R\cdot
                                             f_{\ell 2}^{R*}\\
& & \mbox{\hspace*{-2cm}} {\rm and}\nonumber\\
\Delta_Z(\bar{s}) & = & \frac{g^2}{\cos^2\theta_W}\cdot
\frac{1}{\bar{s}-m_Z^2+i m_Z \Gamma_Z}
\label{5C1}\\
\Delta_L(\bar{t}) & = & \frac{g^2}{\bar{t}-m^2_{\tilde{e}_{L}}}\cdot
\bar{f}_{\ell2}^{L*}\cdot \bar{f}_{\ell 1}^L \label{5C2}
\quad\quad,\quad\quad
\Delta_R(\bar{t}) = \frac{g^2}{\bar{t}-m^2_{\tilde{e}_R}}\cdot
\bar{f}^{R*}_{\ell 2}\cdot \bar{f}^R_{\ell 1} \label{5C3}\\
\Delta_L(\bar{u}) & = & \frac{g^2}{\bar{u}-m^2_{\tilde{e}_L}}\cdot
\bar{f}^L_{\ell 2}\cdot \bar{f}^{L*}_{\ell 1} \label{5C4}
\quad\quad,\quad\quad
\Delta_R(\bar{u}) = \frac{g^2}{\bar{u}-m^2_{\tilde{e}_R}}\cdot
\bar{f}^R_{\ell 2}\cdot \bar{f}^{R*}_{\ell 1} \label{5C5}.
\end{eqnarray}
\section{Numerical Results and Discussion}
\subsection{Scenarios}
Neutralinos are linear superpositions of the photino $\tilde{\gamma}$,
the Zino $\tilde{Z}$ and the two higgsinos $\tilde{H}^0_a$ and
$\tilde{H}^0_b$. The $\tilde{\gamma}$ and $\tilde{Z}$ components
 only couple to the selectrons
 whereas the higgsino components couple to the $Z^0$.
The composition and the masses of the neutralino states depend on the three
SUSY mass parameters $M, M'$ (sometimes also called $M_2$ and $M_1$)
 and $\mu$, whose values follow from the specific SUSY breaking
mechanism, and on the ratio $\tan\beta=v_2/v_1$ of the vacuum expectation
values of the Higgs fields. In order to reduce the number of parameters we
shall assume $M'=\frac{5}{3} M \tan^2 \theta_W$ as suggested by grand
unification \cite{Haber-Kane}. The gaugino mass parameter $M$
is related to the gluino mass by
$M\approx 0.3 m_{\tilde{g}}$ \cite{Bartl} and the gluino mass is roughly given
by $m_{\tilde{g}}\approx 2.4 m_{1/2}$ \cite{LHC}, where
$m_{1/2}$ is the common
 gaugino mass at $M_{GUT}$ ($M\approx 0.72 m_{1/2}$).

The masses of the sleptons are related to the SUSY parameters $M$ and
 $\tan\beta$ and to the common
scalar mass parameter $m_0$ at the unification point
\cite{Bartl}:
\begin{eqnarray}
& & m_{\tilde{\ell}_L}^2=m_0^2+0.79 M^2 +m_z^2 \cos 2\beta(-0.5+\sin^2\theta_W)\\
& & m_{\tilde{\ell}_R}^2=m_0^2+0.23 M^2 -m_z^2 \cos 2\beta \sin^2\theta_W.
\end{eqnarray}

In order to illustrate the influence of the neutralino mixing and
 of the scalar mass $m_0$
we shall consider three representative scenarios which differ significantly
in the nature of the two lowest mass eigenstates $\tilde{\chi}^0_1$
and $\tilde{\chi}^0_2$.
 The selectron masses are calculated for two values of the scalar mass $m_0$,
 $m_0=90$ GeV
and $m_0=200$ GeV.
For the parameters of the Standard Model (SM) we take $m_Z=91.19$ GeV,
$\Gamma_Z=2.49$ GeV, $\sin^2\theta_W=0.23$ \cite{Data} and
$\alpha=1/128$. We choose $\tan\beta=2$.
 The parameters of our scenarios and the mass eigenvalues of the two lightest
neutralinos, the light chargino and the selectrons are given in TABLE~II.
 The width of the $\tilde{\chi}^0_2$ has been computed with the
 program of \cite{Hessel} (see TABLE~II).

Notice that also in scenarios with $\mu<0$ the branching ratio for the
radiative decay is less than $0.5\%$ in the examined region of parameter
space \cite{Mele2}.

The $\tilde{\gamma}$ and $\tilde{Z}$ are mixtures of the $\tilde{B}$
and the $\tilde{W}_3$ gauginos, 
$\tilde{\gamma}=\cos\theta_W \tilde{B}+\sin \theta_W \tilde{W}_3$,
$\tilde{Z}=-\sin\theta_W \tilde{B}+\cos\theta_W \tilde{W}_3$.
 Therefore in TABLE~III the components of the neutralino states are given in 
 the basis ($\tilde{\gamma}, \tilde{Z}, \tilde{H}^0_a, \tilde{H}^0_b$)
and in the basis ($\tilde{B}, \tilde{W}_3, \tilde{H}^0_a, \tilde{H}^0_b$).

In scenario A $\tilde{\chi}^0_1$ has a dominating photino component and
$\tilde{\chi}^0_2$ has a dominating zino component, whereas in
scenario B both neutralinos are nearly equal photino-zino mixtures.
In the two scenarios A and B the $\tilde{\chi}^0_1$
 is almost a pure
B-ino and the neutralino $\tilde{\chi}^0_2$ 
is nearly a pure
$\tilde{W}_3$-ino.
In scenario C both neutralino states are dominated by strong
higgsino components (TABLE~III).

For the phase space integration (chosen relative accuracy $10^{-3}$)
 we used the Monte-Carlo routine Vegas. 
 The evaluation was made for cms-energies of $\sqrt{s}=193$ GeV and
 $\sqrt{s}=500$~GeV.

It can be derived from \cite{Tata} that the total cross section factorizes, 
i.e. it is independent of the spin correlations.
This fact has been used in our numerical calculations as a check
for the phase space integration and the total cross sections are given
in TABLE~IV. 
\subsection{Lepton angular distributions}
In this section we give numerical results for the angular
 distributions of $\ell^{-}$ with respect to the electron beam axis
 computed with complete spin correlations according to 
Eqs.~(52)--(64). The angular 
distributions of $\ell^{+}$   
are obtained by substituting
$\cos\Theta_{M-} \to -\cos\Theta_{M+}$.

In order to demonstrate the significance of the spin correlations
 we compare 
our results
with those obtained from the assumption of factorization of the
differential cross section into production and decay.
As can be seen from FIGS.\ref{fig_4a}--\ref{fig_6b} 
for all mixing scenarios
and both values of the scalar mass $m_0$ the 
contribution of the spin
correlations has the biggest effect in the forward and 
in the backward
direction and vanishes in the direction
perpendicular to the beam axis ($\cos\Theta_{M-}=0$).
The contribution of spin correlations in the forward direction
 has opposite sign of that in the backward direction.
Their magnitude decreases with increasing energy.

Especially for lower energies the spin effect is sizeable 
in scenario A with a
photino-like $\tilde{\chi}^0_1$ and a zino-like 
$\tilde{\chi}^0_2$ for
 both values of $m_0$. For $\sqrt{s}=193$~GeV ($\sqrt{s}=500$~GeV) 
its magnitude amounts to about $20\%$ ($6\%$)
 in the forward and backward direction for both values of $m_0$,
$m_0=90$~GeV and $m_0=200$~GeV, FIG.\ref{fig_4a} (FIG.\ref{fig_4b}).
In both cases the contribution of the spin correlation is 
negative in the 
backward direction and positive in the forward direction.

A comparison with the results for the gaugino-like scenario B
(FIG.\ref{fig_5a} and FIG.\ref{fig_5b})
shows how sensitively the spin correlation
effect depends on the gaugino-higgsino mixing and on the 
value of $m_0$. It is noteworthy that
although in both scenarios A and B the 
$\tilde{\chi}^0_1$ is B-ino-like and
$\tilde{\chi}^0_2$ is $\tilde{W}_3$-ino-like 
with, however, different phases of the $\tilde{W}_3$-ino and 
B-ino admixture. For $\sqrt{s}=193$~GeV and $m_0=90$~GeV
(FIG.\ref{fig_5a}) the 
magnitude of the spin effect is only 2\% in the forward and in 
the backward direction whereas for $m_0=200$~GeV (FIG.\ref{fig_5b}) 
it amounts to 10\%.
For higher energy $\sqrt{s}=500$~GeV it is negligible (FIG.\ref{fig_5c}).

It is remarkable that in the case of gaugino-like neutralinos the
scalar mass $m_0$ crucially determines the 
shape of the angular distributions.
This is most obvious for scenario B and $\sqrt{s}=193$~GeV. 
For the smaller value $m_0=90$ GeV the 
angular distribution has a maximum nearly perpendicular to the beam
direction and is almost FB symmetric, $A_{FB}=-0.8\%$ (compare 
TABLE~V). The contribution of the spin
 correlations is positive in the backward direction 
and negative in the 
forward direction (FIG.\ref{fig_5a}). Increasing the value of $m_0$
 completely changes the shape of the angular distribution and  
for $m_0=200$~GeV it has a minimum in the backward hemisphere
 and the forward direction is favoured, $A_{FB}=+5.9\%$ (FIG.\ref{fig_5b}). 
Now the contribution
 of the spin correlations is negative in the backward direction and 
positive in the forward direction.

In the higgsino-like scenario C both production and decay are
dominated by $Z^0$-exchange. 
Therefore the dependence on $m_0$ is considerably
smaller and we give only numerical results for $m_0=90$~GeV.
Here the contribution of spin correlations is negligible 
and the angular distribution is practically flat for 
$\sqrt{s}=193$~GeV (FIG.\ref{fig_6a})
and FB-symmetric, $A_{FB}=0.07\%$, and for $\sqrt{s}=500$~GeV 
with a minimum perpendicular 
to the beam direction and $A_{FB}=0.11\%$ (FIG.\ref{fig_6b}).

Notice however that a negligible FB-asymmetry is not an 
unequivocal signature
for higgsino-like neutralinos. As can be seen from 
FIG.\ref{fig_5a} for
scenario B with $m_0=90$~GeV the FB-asymmetry may also be small 
 for gaugino-like neutralinos (see TABLE~V). 
From FIGs.\ref{fig_4b}, \ref{fig_5c}
and \ref{fig_6b} we conclude that for higher energies far 
enough from threshold the lepton angular distribution is 
suitable for distinguishing between gaugino-like and 
higgsino-like neutralinos. For $\sqrt{s}=500$~GeV the angular 
distribution is practically FB-symmetric with however a maximum 
for gauginos but a minimum for higgsinos perpendicular to the beam
 direction.
\subsection{The lepton opening angle distribution}
In contrast to the lepton angular distribution,
 the distribution of the opening angle between both leptons
factorizes. Due to the Majorana character of the decaying neutralino 
the spin correlation
 terms are just cancelled by this
partial phase space integration.

It is also noteworthy that in contrast to the lepton 
angular distributions, 
 the distributions
of the lepton opening angle are similar for both
gaugino-like scenarios A and B (FIGs.\ref{fig_7} and \ref{fig_8}).
Especially for $\sqrt{s}=193$~GeV they differ,
however, distinctively from that for higgsino-like neutralinos in
scenario C (FIG.\ref{fig_9}).

In the case of gaugino-like neutralinos the distributions for
$\sqrt{s}=193$~GeV are rather
flat (FIGs.\ref{fig_7} and \ref{fig_8}).
 Changing the scalar mass from
$m_0=90$ GeV to $m_0=200$ GeV  results in a reduction of the
cross sections by a factor of approx. 3 in
scenario A and by a factor of approx. 4 in scenario B. The
shape of the distribution, however, remains essentially unchanged.

One should note that for the shape of the opening angle distributions the
 influence of varying the value of $m_0$ is much smaller
 than for the lepton angular distributions.

For higgsino-like neutralinos (FIG.\ref{fig_9}), 
 the shape of the opening angle distribution for $\sqrt{s}=193$~GeV 
is completely different from 
those of gaugino-like neutralinos.
Here the
lepton pairs are preferably emitted with small angles
between them, approximately $66\%$ of them with
an opening angle between 0 and $\pi/2$.

With increasing energy the opening angle distribution shrinks more and
more and for $\sqrt{s}=500$~GeV it displays for both higgsino- and
gaugino-like neutralinos a rather narrow peak at or near
$\cos\Theta_{+-}=1$. Obviously the distribution of the opening angle
between the leptons is suitable to distinguish between gaugino- and
higgsino-like neutralinos at lower energies.
\subsection{Energy Distributions}
Just as the opening angle distribution the energy distribution
of the outgoing lepton factorizes.
 As a consequence of CP invariance (Sec.~2.1) and the Majorana
character of the neutralinos the energy spectra
 of both leptons, $\ell^{-}$ and $\ell^{+}$, are identical
\cite{Petkov}.

In FIG.\ref{fig_10}
 we give the energy distributions for scenario~B for $m_0=90$ GeV and
$m_0=200$ GeV and cms-energies of $\sqrt{s}=193$ GeV and $\sqrt{s}=500$~GeV.
 For all scenarios the
position of the maximum is independent of the actual value of $m_0$.
\subsection{Summary}
In this paper we have calculated the analytical expression for the
differential cross section of the associated production
 of neutralinos, $e^{-} + e^{+}\to
\tilde{\chi}^0_1 + \tilde{\chi}^0_2$,   
and the subsequent
direct leptonic decay, $\tilde{\chi}^0_2 \to 
\tilde{\chi}^0_1 +\ell^{+} + \ell^{-}$, 
 with complete spin correlations between production and decay.
The angular and the energy distribution of the outgoing  lepton
 as well as the distribution of the opening angle between
 both leptons have been computed for cms-energies of $\sqrt{s}=193$ GeV
 and $\sqrt{s}=500$~GeV.
These distributions have been examined with regard to their dependence on
spin correlations, on the neutralino mixing character and on the
scalar mass parameter $m_0$.

The quantum mechanical interference terms between the various
polarization states of the decaying neutralino give rise to a strong effect
in the lepton angular distribution with respect to the beam axis,
 whereas the opening
angle distribution and the energy distribution are independent from these
spin correlations.

For energies not too far above the threshold ($\sqrt{s}=193$~GeV) 
the opening angle distribution turns out to be suitable for distinguishing
 between higgsino-like and gaugino-like neutralinos. However, it is 
 rather insensible to variable mixing in the gaugino sector. Here the shape
of the opening angle distribution only slightly depends on the scalar mass
 $m_0$.

 The lepton angular distribution, on the other hand, is for lower
energies not only 
 very sensitive to the mixing in the gaugino sector but also to the
actual value of $m_0$.

For energies far above threshold the shape of the lepton angular
distribution is rather insensible to the mixing in the gaugino sector
and to the value of $m_0$. It is, however, extremely different for
gaugino- and higgsino-like neutralinos and suitable for distinguishing
between them.

If the neutralinos are gaugino-like,
 the effect of spin correlations in the angular
distributions can be large
amounting to about 20\%. For higgsino-like neutralinos, on the contrary,
the contribution of
spin correlations is practically negligible.

The energy distributions and the distributions of the opening angle, 
finally, are independent from the spin
correlations. Apart from the magnitude of the cross sections they are
rather insensitive to the actual value of $m_0$.\\

The clear structure of the analytical formulae presented here 
allows to include
hadronic decays and to extend the investigations to cascade decays and
to production and decay of, for instance, $\tilde{\chi}^0_2
\tilde{\chi}^0_2$ pairs. Concerning the hadronic decays, 
$\tilde{\chi}^0_2 \to \tilde{\chi}^0_1 q \bar{q}$, where the outgoing
quarks develop two jets, we expect a similar shape of the opening
angle distribution between quark and antiquark. Thus, the two jets
would be better separated for gaugino-like neutralinos than for
higgsino-like neutralinos which prefer small opening angle.

These investigations as well as the inclusion of beam polarization and
the results for the chargino process angular distributions will be
 discussed in a forthcoming paper with regard to the determination of
SUSY parameters.
\section{Acknowledgement}
We thank A.~Bartl and W.~Majerotto for many useful discussions.
We are grateful to V.~Latussek for his support in the development of the
numerical program. This work was also supported by the 
Deutsche Forschungsgemeinschaft under contract Fr 1064/2-2 and the 
`Fond zur F\"orderung
der wissenschaftlichen Forschung' of Austria, Project No. P10843-PHY.

%\end{thebibliography}
\newpage
\begin{figure}
\caption{Feynman graphs for production and leptonic decay.}\label{fig_1}
\end{figure}

\begin{figure}
\caption{The amplitude squared for the combined process including 
spin correlations is composed by the spin density matrix for the 
production and the decay matrix times the pseudopropagator $\Delta_2$
squared (compare Eq.~(9)).}\label{fig_2}
\end{figure}

\begin{figure}
\caption{Configuration of momenta in the laboratory system. The lepton
angular distribution refers to $\Theta_{M-}$, the opening angle
distribution to $\Theta_{+-}$.}\label{fig_3}
\end{figure}

\begin{figure}
\caption{Lepton angular distribution for $\sqrt{s}=193$~GeV in 
scenario A for $m_0=90$ GeV
with spin correlations fully taken into account (upper solid) and
for assumed factorization (upper dotted);
for $m_0=200$ GeV with spin correlations (lower solid) and for
assumed factorization (lower dotted).}\label{fig_4a}
\end{figure}

\begin{figure}
\caption{Lepton angular distribution for $\sqrt{s}=500$~GeV in
scenario A for $m_0=90$ GeV
with spin correlations fully taken into account (upper solid) and
for assumed factorization (upper dotted);
for $m_0=200$ GeV with spin correlations (lower solid) and for
assumed factorization (lower dotted).}\label{fig_4b}
\end{figure}

\begin{figure}
\caption{Lepton angular distribution for $\sqrt{s}=193$~GeV in
scenario B for $m_0=90$ GeV 
with spin correlations fully taken into account (solid) and for assumed
 factorization (dotted).}\label{fig_5a}
\end{figure}

\begin{figure}
\caption{Lepton angular distribution for $\sqrt{s}=193$~GeV in 
scenario B for $m_0=200$ GeV
with spin correlations fully taken into account (solid)
and for assumed factorization (dotted).}\label{fig_5b}
\end{figure}

\begin{figure}
\caption{Lepton angular distribution for $\sqrt{s}=500$~GeV in
scenario B for $m_0=90$ GeV
with spin correlations fully taken into account (upper solid) and
for assumed factorization (upper dotted);
for $m_0=200$ GeV with spin correlations (lower solid) and for
assumed factorization (lower dotted).}\label{fig_5c}
\end{figure}

\begin{figure}
\caption{Lepton angular distribution for $\sqrt{s}=193$~GeV in 
scenario C for $m_0=90$ GeV 
with spin correlations fully taken into account (solid)
and for assumed factorization (dotted).}\label{fig_6a} 
\end{figure}

\begin{figure}
\caption{Lepton angular distribution for $\sqrt{s}=500$~GeV in 
scenario C for $m_0=90$ GeV 
with spin correlations fully taken into account (solid)
and for assumed factorization (dotted).}\label{fig_6b} 
\end{figure}

\begin{figure}
\caption{Opening angle distribution in scenario A for
  $\sqrt{s}=193$~GeV and $m_0=90$ GeV (upper solid)
and $m_0=200$ GeV (lower solid);
for $\sqrt{s}=500$~GeV and $m_0=90$ GeV (upper dotted)
and $m_0=200$ GeV (lower dotted);.}
\label{fig_7}\end{figure}

\begin{figure}
\caption{Opening angle distribution in scenario B for 
 $\sqrt{s}=193$~GeV and $m_0=90$ GeV (upper solid)
and $m_0=200$~GeV (lower solid);
for $\sqrt{s}=500$~GeV and $m_0=90$ GeV (upper dotted) 
and $m_0=200$~GeV (lower dotted).}\label{fig_8}
\end{figure}

\begin{figure}
\caption{Opening angle distribution in scenario C for
  $\sqrt{s}=193$~GeV and $m_0=90$ GeV (solid);
 for $\sqrt{s}=500$~GeV and $m_0=90$~GeV (dotted).}\label{fig_9}
\end{figure}

\begin{figure}
\caption{Energy distribution in scenario B for $\sqrt{s}=193$~GeV and 
$m_0=90$ GeV (upper solid) and $m_0=200$ GeV (lower solid);
for $\sqrt{s}=500$~GeV and 
$m_0=90$ GeV (upper dotted) and $m_0=200$ GeV (lower dotted).}\label{fig_10}
\end{figure}

\newpage
\mediumtext
\begin{table}[h]
\begin{center}
\begin{tabular}{|c||c|c|c|c|c|c|c|}
 Production & $\mu$ & $\nu$ & $\omega$ & $\tau$ & $\upsilon$ &
Couplings & Momenta \quad\\ \hline
& & & & & & &  \\
$\tilde{e}_L \to \tilde{e}_R$ &+1 & $-1$ &$-1$ & $+1$ & $ -1$
& $O^L\leftrightarrow O^R$, $L_e\leftrightarrow R_e$ & \\ \hline
& & & & & & & \\
 $t \to u$ & $-1$ &+1 &$-1$ & $+1$ & $-1$ & $O^L \leftrightarrow O^R$ &
 $k_1 \leftrightarrow k_2$ \\ \hline
decay &  \multicolumn{7}{c|}{} \\ \hline & & & & & & & \\
 $\tilde{\ell}_L\to\tilde{\ell}_R$ &+1 &+1 &+1 & $+1$ & $-1$ &
$\bar{O}^L\leftrightarrow \bar{O}^R$, $\bar{L}_e\leftrightarrow\bar{R}_e$ & \\
\hline & & & & & & & \\
$\bar{t}\to \bar{u}$ &+1 &+1 & $+1$ & $-1$ & $-1$ &
$\bar{O}^L\leftrightarrow \bar{O}^R$ & $p_2 \leftrightarrow p_3$
\end{tabular}

\vspace{.5cm}
\caption{Substitution rules for $W_{ab,cd}$, see Sec.~3.}
\end{center}\label{tab1}
\end{table}

\mediumtext
\begin{table}[h]
\begin{center}
\begin{tabular}{|l||c|c|c|c|c|c|c|c|c|c|c|c|}
\hspace{.5cm} & tan$\beta$ & $ M$ & $ \mu$ & $ \eta_1 m_1$ &
$ \eta_2 m_2$ &  $\eta_{\tilde{\chi}_1^{+}} m_{\tilde{\chi}_1^{+}}$ & 
$ m^{90}_{\tilde{\ell}_{L}}$ &
 $ m^{90}_{\tilde{\ell}_{R}}$ &  $ m^{200}_{\tilde{\ell}_{L}}$ &
$ m^{200}_{\tilde{\ell}_{R}}$ & $\Gamma^{90}_{\tilde{\chi}^0_2}$ & 
$\Gamma^{200}_{\tilde{\chi}^0_2}$\\ \hline
A & $ 2$ & $ 84$ & $ -250$ & $ 46$ & $ 97$ & $97$ & $123$ & $104$ & $217$ &
$207$ & 35.5 & 1.33  \\ \hline
B & $ 2$ &  $ 112$ & $ 448$ & $ 51$ & $ 98$ & $-97$ & $139$ & $110$ & $226$ &
$210$ & 15.0 & 0.74 \\ \hline
C & $ 2$ &  $ 215$ & $ -83$ & $ 76$ & $ -109$ & $97$ & $214$ & $114$ & $279$ &
$228$ & 23.6 & 23.7
\end{tabular}
\end{center}

\vspace{.5cm}
\caption{Parameters $M$, $\mu$ and mass eigenvalues
 in GeV, total width of $\tilde{\chi}^0_2$ in keV. The superscripts denote the
value of the scalar mass $m_0$.}
\end{table}

\narrowtext
\begin{table}[h]
\begin{center}
\begin{tabular}{|l||c|c|}
\hspace{.5cm} &
 $ \tilde{\chi}^0_1$  & $ \tilde{\chi}^0_2$  \\ \hline & &\\
 & $ (\hspace{.2cm} \tilde{\gamma}\hspace{.2cm}|\hspace{.2cm} 
\tilde{Z} 
\hspace{.2cm}|\hspace{.2cm} \tilde{H}^0_a\hspace{.2cm} |\hspace{.2cm}
 \tilde{H}^0_b)$\hspace{.2cm} 
 & $ (\hspace{.2cm}\tilde{\gamma}\hspace{.2cm}| \hspace{.2cm}
\tilde{Z}\hspace{.2cm} | \hspace{.2cm}\tilde{H}^0_a \hspace{.2cm}|
\hspace{.2cm} \tilde{H}^0_b\hspace{.2cm})$\\ \hline
A & $
(+.94\mid-.33\mid-.08\mid-.08)$
& $ (-.35\mid-.89\mid-.16\mid-.23)$ \\ \hline
B & $
(+.79\mid-.60\mid+.11\mid+.07)$
& $ (-.62\mid-.76\mid+.17\mid+.10)$ \\ \hline
C & $
(-.17\mid+.22\mid-.19\mid+.94)$
& $ (-.05\mid+.29\mid-.92\mid-.26)$ \\ \hline & & \\
 & $ (\hspace{.2cm} \tilde{B}\hspace{.2cm}|\hspace{.2cm}
 \tilde{W}_3
\hspace{.2cm}|\hspace{.2cm} \tilde{H}^0_a\hspace{.2cm}|\hspace{.2cm}
\tilde{H}^0_b\hspace{.2cm})$ 
& $ (\hspace{.2cm} \tilde{B}\hspace{.2cm}|\hspace{.2cm}
 \tilde{W}_3 \hspace{.2cm}|\hspace{.2cm} 
\tilde{H}^0_a\hspace{.2cm}|\hspace{.2cm} \tilde{H}^0_b\hspace{.2cm})$
\\ \hline
A & $ (+.98\mid+.16\mid-.08\mid-.07)$ & $ (+.12\mid-.95\mid-.16\mid-.23)$
\\ \hline
B & $ (+.98\mid-.15\mid-.13\mid+.09)$ & $ (-.18\mid-.96\mid-.18\mid-.10)$\\
\hline
C & $ (-.25\mid+.11\mid-.19\mid+.96)$ & $ (-.18\mid+.23\mid+.92\mid+.24)$
\end{tabular}
\end{center}

\vspace{.5cm}
\caption{Neutralino eigenstates.}
\end{table}

\mediumtext
\begin{table}[h]
\begin{center}
\begin{tabular}{|l||c|c||c|c|}
\multicolumn{5}{|c|}{$\sigma(e^- e^+ \to \tilde{\chi}^0_1
\tilde{\chi}^0_2 \to \tilde{\chi}^0_1 \tilde{\chi}^0_1 \ell^+ \ell^-) \quad
/fb$}\\
 \hline
\hspace{.5cm} & \multicolumn{2}{c||}{$\sqrt{s}=193$~GeV} &
\multicolumn{2}{c|}{$\sqrt{s}=500$~GeV}\\ \hline
\hspace{.5cm} & $m_0=90$~GeV &
 $m_0=200$~GeV & $m_0=90$~GeV & $m_0=200$~GeV \\ \hline
A & 38.9 & 11.1 & 32.7 & 21.6 \\ \hline
B & 10.0 & 2.5 & 10.9 & 6.1\\ \hline
C & 23.3 & 23.4 & 6.0 & 6.1
\end{tabular}
\end{center}

\vspace{.5cm}
\caption{Total cross sections for $e^- e^+ \to \tilde{\chi}^0_1
\tilde{\chi}^0_2$ and subsequent leptonic decay, 
$\tilde{\chi}^0_2 \to
\tilde{\chi}^0_1 \ell^+ \ell^-$ for $\sqrt{s}=193$~GeV and
$\sqrt{s}=500$~GeV with $m_0=90$~GeV and $m_0=200$~GeV.}
\end{table}

\mediumtext
\begin{table}[h]
\begin{center}
\begin{tabular}{|l||c|c||c|c|}
\multicolumn{5}{|c|}{
$A_{FB}=[\sigma(\cos\Theta_{-}>0)-\sigma(\cos\Theta_{-}<0)]
/[\sigma(\cos\Theta_{-}>0)+\sigma(\cos\Theta_{-}<0)]
 \quad /[\%]$}\\
 \hline
 & \multicolumn{2}{c||}{$\sqrt{s}=193$~GeV} &
\multicolumn{2}{c|}{$\sqrt{s}=500$~GeV}\\ \hline
\hspace{.5cm} & $m_0=90$~GeV &
 $m_0=200$~GeV & $m_0=90$~GeV & $m_0=200$~GeV \\ \hline
A & 10.2 & 11.6 & 2.3 & 3.0 \\ \hline
B & $-0.8$ & 5.9 & $-0.3$ & 1.8\\ \hline
C &  0.07 & 0.11 & $-0.01$ & 0.00
\end{tabular}
\end{center}

\vspace{.5cm}
\caption{Forward-Backward-Asymmetry $A_{FB}$ for 
$\sqrt{s}=193$~GeV 
and $\sqrt{s}=500$~GeV with $m_0=90$~GeV and $m_0=200$~GeV.}
\end{table}
\end{document}